\documentclass[a4paper,11pt]{article}
\usepackage{jinstpub} 
\usepackage{lineno}
\linenumbers

\title{SBC-SNOLAB scintillation system and SiPM implementation for dark matter searches}

\collaboration[c]{on behalf of the SBC collaboration}

\author{H. Hawley-Herrera}

\affiliation{Queen`s University, Department of Physics, Engineering Physics, and Astronomy, Kingston, K7L 3N6, Canada}
\emailAdd{hawleyherrera.h@queensu.ca}

\abstract{The Scintillating Bubble Chamber (SBC) collaboration is constructing a 10~kg liquid argon (LAr) bubble chamber at SNOLAB called SBC-SNOLAB having the main objective of detecting dark matter. One of the most novel aspects of SBC-SNOLAB is the scintillation system, consisting of LAr doped with on the order of 10~ppm Xe, 48 FBK VUV silicon photomultipliers (SiPMs), the SiPM electronics, two quartz jars, and liquid CF$_4$ used as an hydraulic fluid and additional source of scintillation photons. In contrast with traditional single or dual phase scintillation experiments, the collected LAr scitillation light is used to veto signals which involve the detection of at least a single photoelectron. These proceedings will describe in detail the current SBC-SNOLAB scintillation system which includes the unique design considerations for SBC-SNOLAB that limit the light collection efficiency and the electronics.}


\begin{document}
\nolinenumbers
\maketitle
\flushbottom

\section{Introduction}
The goal of the Scintillating Bubble Chamber (SBC) collaboration is to detect dark matter utilizing the well-demonstrated technology of bubble chambers \cite{cite:pico60} with scintillators as the active fluid. The main benefit scintillators provide to a bubble chamber is the additional channel for tagging background events and performing energy reconstruction. Examples of scintillators that can be used for bubble chambers are noble elements, liquid nitrogen, and CF$_4$. If the scintillator is a liquid noble element, there is an intrinsic reduction of the energy threshold as demonstrated in ref.~\cite{cite:lilXe}.

The SBC detector called SBC-SNOLAB is a 10~kg liquid argon (LAr) bubble chamber to be built at the underground physics laboratory, SNOLAB\footnote{\url{https://www.snolab.ca/about/about-snolab/}}. Its construction design follows the same buffer free-liquid design as PICO-40L\footnote{\url{https://www.picoexperiment.com/pico-40l/}} bubble chamber as demonstrated in ref.~\cite{cite:pico40L}. SBC-SNOLAB is a scaled up version of the smaller 30~g xenon bubble chamber built in Northwestern University where the fundamental detector operating principles were first tested \cite{cite:lilXe}.

SBC-SNOLAB consists of four essential systems. One, the thermo-mechanical system, has the objective of starting the event cycle by pressurization and setting the temperature so a superheated state is achieved; stopping the bubble formation by compressing the active fluid; creating a temperature gradient that stops undesired components of the active fluid from nucleating; and setting the energy threshold. Second is the camera system, consisting of three cameras that record the expansion of the bubble after a nucleation. Third is the piezoelectric system consisting of several piezos located around the active fluid listening for the bubble formation. Lastly, the scintillation system collects the photons created during an event. For more information about other aspects of the chamber and physics goals not covered in these proceedings see refs.~\cite{cite:white_paper, cite:universe}.

These proceedings will describe the current design for the scintillation system which is the most novel component of SBC-SNOLAB. Discussion of the importance of the scintillation system for the search for dark matter will be included with comparisons to other scintillation-based particle detectors. This discussion will also include a unique set of problems which come associated with low background detectors regarding the design and use of the silicon photomultipliers (SiPMs), electronics and acquisition system.

\section{The scintillation system}
\begin{figure}
    \centering
    \includegraphics[width=0.85\textwidth]{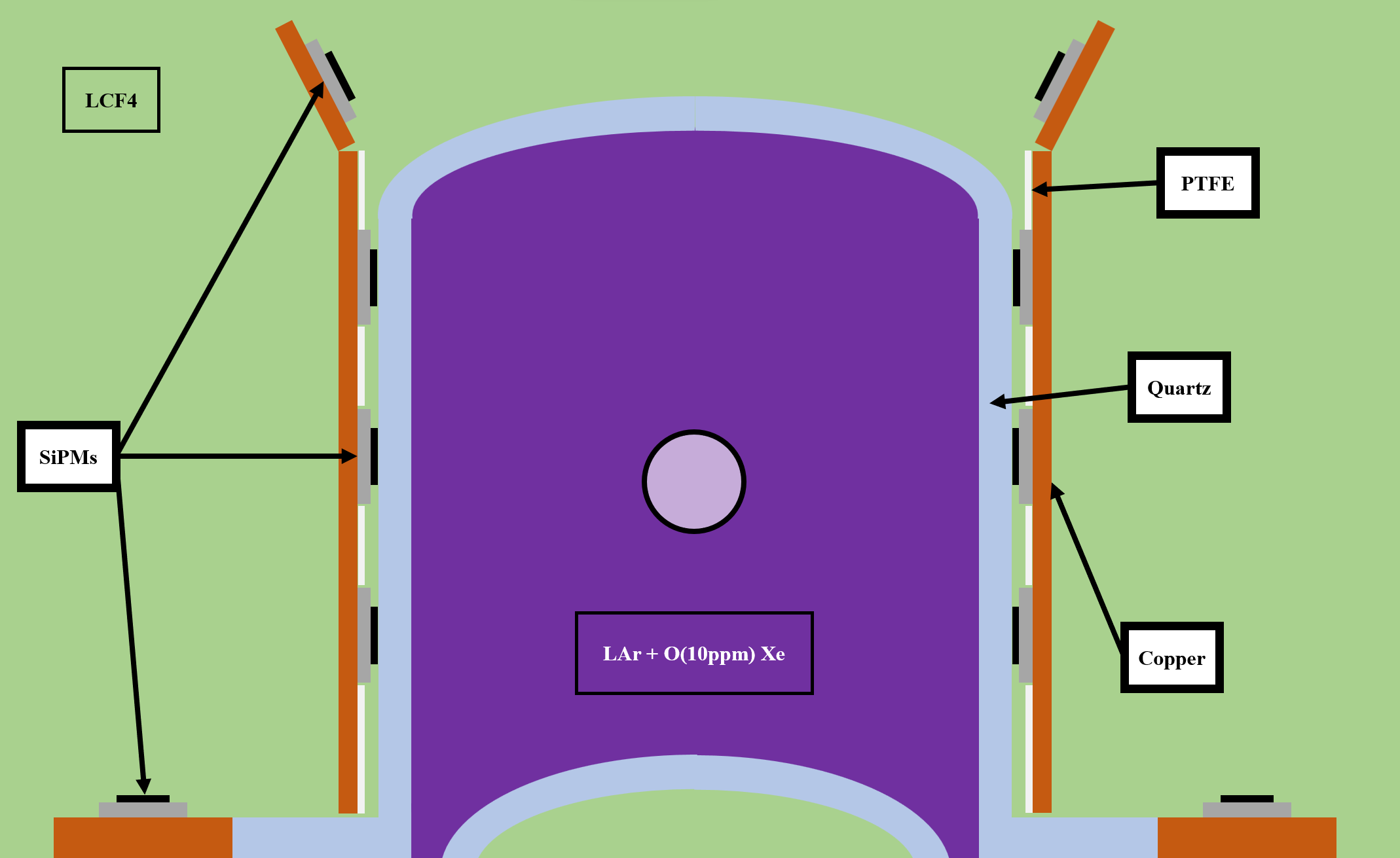}
    \caption{Simplified diagram of SBC-SNOLAB scintillation system. It consists of a liquid argon (LAr) doped with on the order of 10~ppm Xe located between two quartz jars using "right side up" orientation of PICO-40L construction \cite{cite:pico40L}. The light collection devices consists of 32 FBK vacuum ultraviolet (VUV) SiPMs collecting the LAr scintillation and 16 SiPMs are used to collect the LCF$_4$ scintillation. Additionally, thin polytetrafluoroethylene (PTFE) sheets are used to cover the side of the outer jar to reflect the LAr scintillation.}
    \label{fig:scintillation_system}
\end{figure}

The SBC-SNOLAB scintillation system is composed primarily of 10~kg of LAr located in between two concentric quartz jars (constructed of Hereaus Suprasil 310\footnote{\url{https://www.heraeus-group.com/en/}}) and 32 SiPMs surrounding the outer-most quartz jar. Additionally, the SiPMs and the quartz jars are submerged in a bath of liquid CF$_4$ that acts as a thermo-mechanical exchange fluid and, as will be discussed later, as an additional source of data which can be used for background suppression. A simplified schematic diagram of the SBC-SNOLAB scintillation system can be seen in figure~\ref{fig:scintillation_system}. Each component of the scintillation system will be described in detail in this section with the justifications behind some of the unconventional design decisions.

\subsection{LAr}
In SBC-SNOLAB, LAr was chosen as the active fluid because of the expected increased sensitivity to lower dark matter masses. LAr emits 128~nm scintillation photons with a yield approximately of $40\,000$ photons per MeV for ionizing radiation \cite{cite:lar_properties}. The scintillation is emitted following the interaction under two time constants: the singlet state with a decay constant of a few nanoseconds, and the triplet state in the single microsecond range. However, if pure LAr is used, the quartz jars will absorb most if not all of the LAr scintillation light as expected from ref.~\cite{cite:quartz}. 

A simple but naive solution is to place the SiPMs in between the quartz jars with the LAr, but sharp features (corners, the silicon circuits, and surface roughness) create nucleation points which would likely increase the nucleation to rates beyond the maximum limit of approximately $1\,000$ events per day. An alternative solution which avoids the excessive nucleation sites is to waveshift the scintillation light to wavelengths which allow for transmission through the quartz. 

The SBC collaboration has chosen Xe as the SBC-SNOLAB wavelength shifter as it has been shown to convert the 128~nm scintillation of Ar to the Xe scintillation wavelength of 174~nm \cite{cite:xe_doping_one}. Additionally, the Xe scintillation has also been shown to be almost fully reflective in polytetrafluoroethylene (PTFE) \cite{cite:ptfe_reflectivity} which can be used to increase the light collection slightly by covering the quartz in PTFE sheets. The light collection efficiency also improves because Xe increases the scintillation photons yield \cite{cite:xe_doping_two}. 


\subsection{Quartz jars}
Quartz was chosen as the jar material for its radiopurity and visible-wavelength transparency and smooth inner surface (to prevent nucleations). The inner jar acts as a piston to control the pressure of the LAr while the outer one remains static. Both jars are sealed to a stainless steel bellows with spring-energized PTFE seals. 

\begin{figure}
    \centering
    \includegraphics[width=0.6\textwidth]{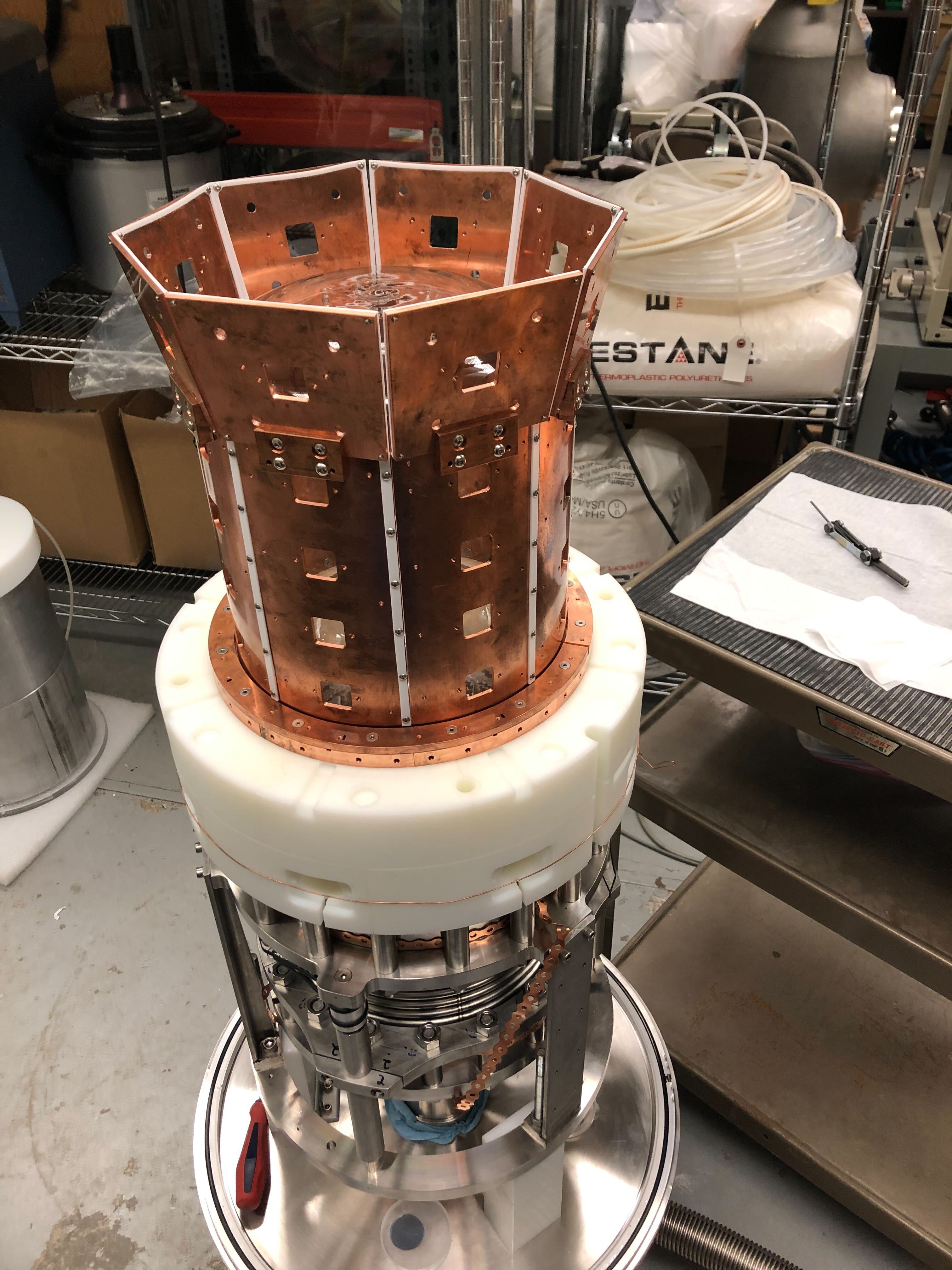}
    \caption{Queen's University test setup is a replica of SBC inner assembly adapted to test the cryogenic feasibility of the quartz jars seals and the SiPM acquisition chain. No LAr or LCF$_4$ are found in this chamber.}
    \label{fig:test_setup}
\end{figure}

A test setup of the quartz jars has been replicated at Queen's University with the goal of testing the cryogenic feasibility of the seals and jars. They were successfully tested by pulling a high vacuum on the outside of the jars while the steel frame is thermally connected to a cryohead using several strands of thick copper braided wire. The test setup can be seen in figure~\ref{fig:test_setup}.

\subsection{SiPMs}
The SiPMs play an important role in the scintillation system collection efficiency via the photon detection efficiency, and the fill factor. Currently, for radio-purity reasons, SBC-SNOLAB is planning to use up to 48 Fondazione Bruno Kessler (FBK) vacuum ultraviolet (VUV) SiPMs a variation of the SiPMs discussed in ref.~\cite{cite:fbk_vuv_sipms}. 32 SiPMs are evenly distributed facing the quartz jars and the remaining 16 are used to look outwards into the CF$_4$ space. They are connected to a custom made TRIUMF-built amplifier outside the chamber using 5~m to 7~m long coaxial cables that also carry the power to the SiPMs. The SiPMs are mechanically attached to the copper panels with a thin layer of liquid CF$_4$ between them and the jars and no optical coupling material in between. As of this moment, no material has been found to be compatible with LCF$_4$ that also has an index of refraction equal to that of the quartz jars.

An indirect impact of the SiPMs comes in the form of delayed correlated avalanches, DCAs. They are created when an independent avalanche (due to an incident photon or generated thermally) or other correlated avalanches start a series of mechanisms in the silicon and construction materials of the SiPMs that start another avalanche within the SiPM \cite{cite:sipm_basics}. Understanding the probability of DCAs and their time distribution is of importance for SBC-SNOLAB as it will impact the acquisition window and will define the pre-acquisition window required to minimize bubbles coincident with DCAs being misinterpreted as bubbles with scintillation. Currently, SBC is preparing a paper for the characterization of the SiPMs to understand DCAs and to set up the SiPMs to maximize their signal-to-noise ratio.

Another important component of the scintillation system that is not found inside the SBC-SNOLAB chamber are the SiPM acquisition electronics. They are used to amplify the SiPM pulses, filter noise, and save the data to disk. The data acquisition electronics consist of the custom-made TRIUMF amplifiers, the coaxial cables, a CAEN 1740D 64 channel 62.5~MS/s digitizer, and computer for storage.

The amplifiers are found outside the SBC-SNOLAB pressure chamber or in air space as most electronic components such as resistors, capacitors, and connectors are too radioactive to be placed near the detector \cite{cite:electronics_background_legend, cite:electronics_background_majorana}. As a consequence, long coaxial cables connect the SiPM to the amplifier where noise spurs can be bigger than the signal. Therefore, a hardware low-pass filter is used to reduce the external noise. Signal bandwidth is not important for SBC-SNOLAB because no additional information is gained from pulse timing reconstruction which experiments like DarkSide-50 or XENONnT use to distinguish backgrounds from signal \cite{cite:darkside, cite:xenonNT}. However, a possible long-term solution to the noise problem without sacrificing timing information is the creation of a single device or system-on-a-chip that contains both the SiPM and the processing electronics on a single or multiple silicon chips similar to ref.~\cite{cite:electronics_background_legend}.

The acquisition back-end must be flexible during data taking because SBC-SNOLAB has the potential to do physics beyond dark matter searches which require different acquisition setups to achieve. An example of potential physics is the effects of different levels of Xe dopant or LAr scintillation as a function of temperature and pressure. The data required for the scintillation physics can be acquired during recompressions in when there is no dark matter sensitivity.

\subsection{LCF$_4$}
The original role of the LCF$_4$ is to act a thermo-mechanical exchange fluid which remains stable at LAr temperatures and pressures. However, during an initial R\&D phase, the LCF$_4$ was unexpectedly found to scintillate. The decision was made to place of several SiPMs to collect the scintillation and it is planned to be used as a veto for external sources of backgrounds such as muon-induced neutrons, and gammas. The scintillation of the gaseous CF$_4$ has been documented in \cite{cite:cf4} but no information is available for the liquid state. SBC is currently preparing a paper on the liquid scintillation properties of LCF$_4$.

Nevertheless, the light collection efficiency is important for a dark mater search as it impacts the energy threshold for nuclear recoils, but it is not a priority for the scintillation system. The engineering constrains previously discussed (such as the high number of optical interfaces and low background requirements) of the scintillation system does not favor high collection efficiencies. Nevertheless, the main objective of the scintillation system is to use the collected photons as a veto instead of to perform energy reconstruction, which would be  similar to single or dual phase scintillation experiments. A low and uncertain light collection efficiency will not impact the dark matter search significantly as most of the backgrounds are expected to emit considerable amount of photons. Ideally, the veto scenario would require at least one single photo-electron detected across all SiPMs, but this scenario requires the scintillation-generating backgrounds to be below kHz levels.


\section{Summary}
The SBC collaboration is building SBC-SNOLAB a 10~kg LAr doped with Xe bubble chamber in order to attempt to detect dark matter. The combination of bubble chamber technology with scintillation at cryogenic temperatures, and the additional requirement of low backgrounds, brings a new set of challenges that SBC is undertaking. The material selection is limited to proven low-background materials and any relative radioactive materials have to be placed away from the detector. The light collection efficiency is directly affected by these design constrains. However, the SBC-SNOLAB scintillation system is used as a background veto which does not require high light collection efficiencies which would only impact the efficacy at low energies. Solutions to these problems require long R\&D campaigns, to be completed when SBC is designing bigger chambers.

Currently, the SBC collaboration has finished characterizing the SiPMs and preparing a publication on the measured gains, breakdown voltages, dark noise rates and probabilities of a correlated avalanche. 

Modeling of the optical propagation in SBC-SNOLAB has also started in Geant4\footnote{\url{https://geant4.web.cern.ch/}} which includes the SiPMs, the quartz jars, and the scintillation light. 

The LCF$_4$ has its scintillation properties measured and comparison with a MC model is currently ongoing for publication. Finally, SBC is confident to reach sufficient scintillation collection efficiency for SBC-SNOLAB dark matter goal. 

\bibliographystyle{JHEP}
\bibliography{biblio.bib}
\end{document}